# The safe operating space for greenhouse gas emissions

Steffen Petersen[1*], Morten W. Ryberg[2], Morten Birkved[3]

[1]Department of Civil and Architectural Engineering, Aarhus University, Inge Lehmanns Gade 10, 8000 Aarhus C, Denmark

[2]SWECO, Ørstad Boulevard 41, 2300 Copenhagen S, Denmark

[3]Department of Green Technology, Southern University of Denmark, Campusvej 55, 5230 Odense M, Denmark

*Corresponding author. Phone: +4541893347. Email: stp@cae.au.dk

**Abstract**

The Planetary Boundary for 'Climate Change' has been surpassed, and humanity must therefore decide on a pathway back to the safe operating space below the Planetary boundaries to minimise the risk of deleterious or even catastrophic environmental change at continental to global scales. However, the control variables used in the concept do not link to current indicators quantified in life-cycle assessments. Translating the Planetary Boundaries into maximum scores for life-cycle assessment indicators is therefore necessary to enable the assessment of whether life-cycle environmental impacts of e.g. products and services respect the Planetary Boundaries. This paper investigates two different approaches for translating the Planetary Boundary for 'Climate Change' into estimations of maximum emission of greenhouse gasses expressed as annual emissions of $CO_2$-equivalents – a commonly used 'Climate Change' indicator in life-cycle assessments. The estimations may serve as an end goal benchmark for the life-cycle environmental impact of products and services in proposed pathways and roadmaps for the regeneration back to the Planetary Boundary for 'Climate Change'.

*Note: The paper has not been peer-reviewed and may therefore be subject to changes.*

## 1 Introduction

The concept of Planetary Boundaries expresses a safe operating space for the human impact on Earth's ecosystems [1]. The Earth system *Climate Change* concerns the energy imbalance at the top of Earth's atmosphere (radiative forcing) which is affected by human emissions of greenhouse gasses. The Planetary Boundary for this system is 1 W/m$^2$ which corresponds to a temperature increase of approx. 1 °C relative to the pre-industrial period (the year 1850-1900). This temperature increase was surpassed in 2017 [2] which means that humanity has entered a zone of risk of deleterious or even catastrophic environmental change at continental to global scales [1] – a risk that is increasing due to the continuous increase of human-induced emissions of greenhouse gasses to the atmosphere. Due to the regenerative properties of Earth's ecosystems, it is in theory possible for humankind to return to the safe operating space, but it requires that greenhouse gas emissions are reduced to a level where regeneration is possible. The time it takes for regeneration back to the Planetary Boundary for *Climate Change* depends on the magnitude and rate of greenhouse gas emission reductions – which then again determines the magnitude of the risk for deleterious or catastrophic climate change during the regeneration period. Choosing the pathway back to the safe operating space is thus a political risk-based decision where science only can be used to elucidate the potential consequences of the choice.

The control variables used in the Planetary Boundary concept do not link to current indicators quantified in life-cycle assessments. Translating the Planetary Boundaries into maximum scores for life-cycle assessment indicators is therefore necessary if life-cycle assessments are to support the development of e.g. products and services that respect the Planetary Boundaries. This paper present and applies two different methods for translating the Planetary Boundary of 1 W/m$^2$ to estimations of the maximum annual global emission of greenhouse gases expressed in $CO_2$-equivalents[1]. The estimations express the safe operating space for human-induced emissions of $CO_2$-equivalents assuming that the Planetary Boundary for 'Climate Change' was not surpassed. The estimations therefore only apply as an emission budget once humanity has re-entered the safe operating space below the Planetary Boundaries. At present, the estimations may serve as a benchmark for the end goal for proposed regeneration pathways and roadmaps back to the Planetary Boundary for Climate Change.

## 2  The concept of planetary boundaries

We are currently living in a transitional period between the Holocene and Anthropocene epochs of Earth's history. The Holocene epoch was the interglacial period that began approximately 11 500 years ago that provided a millennial-scale climate cycle [4] fit for Homo Sapiens to create complex societies by modifying and exploiting the environment with ever-increasing scope, from hunter-gatherer and agricultural communities to the highly technological, rapidly urbanizing societies of the 21st century [5]. The new epoch, the Anthropocene, is where humans constitute the dominant driver of change in the Earth System [6]. The shift in epoch has raised the concern that the exponential growth of human-induced pressure on critical biophysical systems may trigger irreversible environmental changes that are potentially catastrophic for humankind. Rocktröm et al. [1] consequentially ask:

*"What are the non-negotiable planetary preconditions that humanity needs to respect to avoid the risk of deleterious or even catastrophic environmental change at continental to global scales?"*

Using the stability of the Holocene state as the scientific reference point for a desirable planetary state, Rockström et al. (ibid.) developed the concept of Planetary Boundaries. The Planetary boundaries define and quantify boundary impact levels for nine Earth System processes (Figure 1) that are key for maintaining the stability of the Earth System and which should not be transgressed[2].

---

[1] $CO_2$-equivalents is a measure where the global warming potential of a greenhouse gas other than $CO_2$ (e.g. methane or nitrous oxide) is normalised relative to the global warming potential of $CO_2$. For example, the global warming potential of one kg of methane corresponds to approx. 30 kg of $CO_2$. The normalisation enables the summing of different greenhouse gas emissions to a total emission with the same unit.
[2] The boundaries from the 2009 publication were updated in 2015 by Steffen et al. [5].

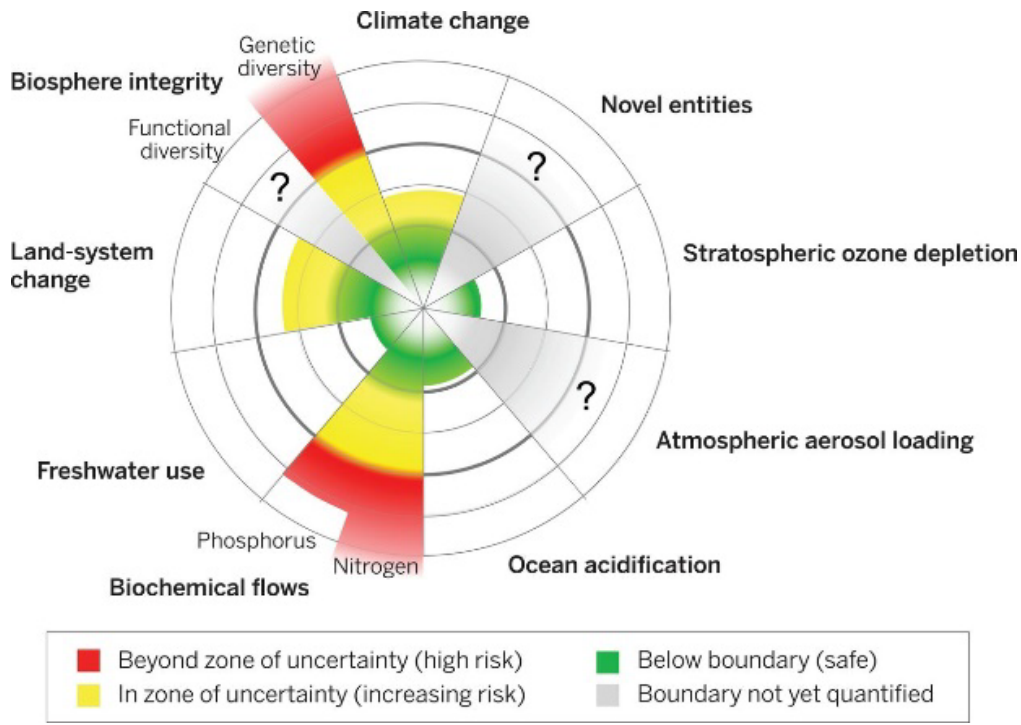

Figure 1: Status of the control variables for seven of the Earth system processes relative to the Planetary boundaries [7].

The Earth System processes in the planetary boundaries can be divided into those operating at a planetary scale and those operating at a sub-planetary scale, see Figure 2.

Indeed, despite the 'Planetary Boundary' naming, only the Earth System processes for climate, ocean acidification, and stratospheric ozone depletion operate at the planetary scale. For instance, the climate system has many known tipping points, that when crossed can initiate non-linear, abrupt, irreversible environmental changes that are potentially catastrophic to the stability of the Holocene state of Earth.

The remaining Earth System processes operate at a sub-planetary scale without known global scale thresholds [1]. Thus, the Planetary Boundaries in the Planetary Boundary framework distinguish between sharp continental or planetary thresholds which can affect the Earth System through systemic top-down processes, and "boundaries based on 'slow' planetary processes with no current evidence of threshold behaviour on a planetary scale, but which provide the underlying resilience of the Earth System by functioning as sinks and sources of carbon and by regulating water, nutrient, and mineral fluxes [1].

The concern for the slow planetary boundaries is that local to regional scale thresholds are exceeded which can become a global concern if these thresholds are exceeded in multiple locations simultaneously or if the gradual aggregate impacts may add up and affect other Earth System processes and, thereby, the general stability of the Earth System through environmental interlinkages and feedback mechanisms [1]. Due to the interlinkages among Earth systems, the Planetary boundaries concept defines "safe operating spaces" for both global and local/regional Earth systems to minimize the risk of destabilizing the Holocene state.

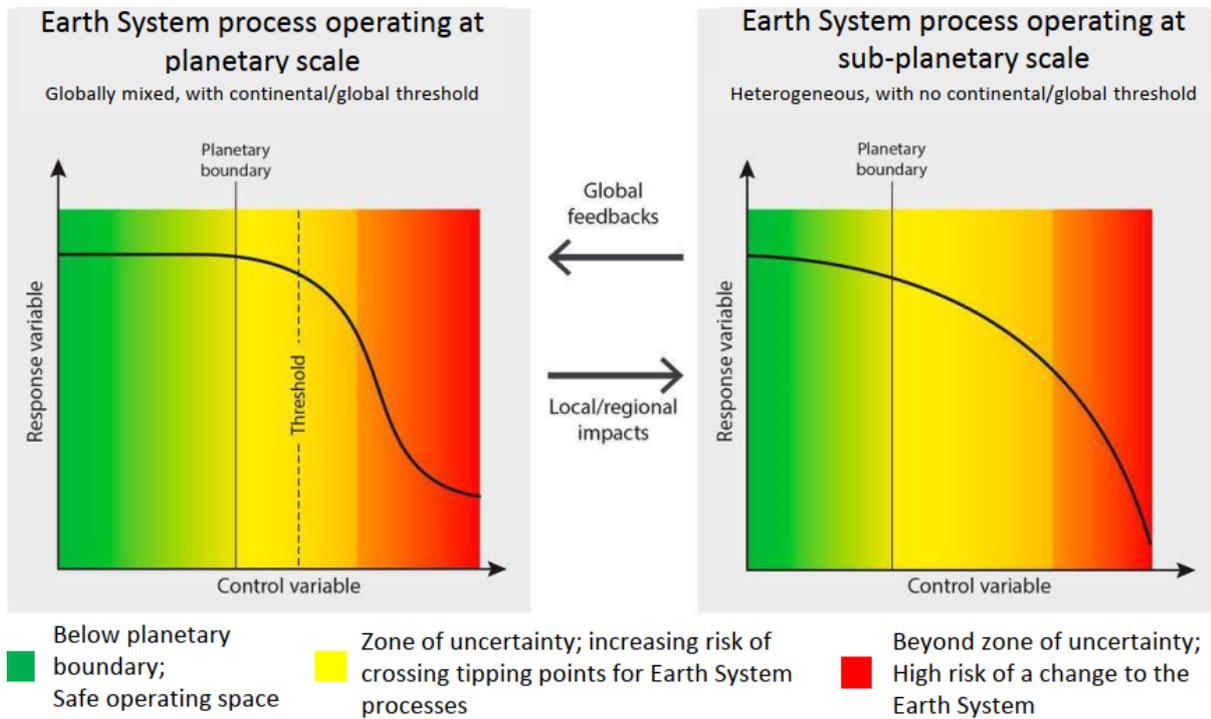

Figure 2: The conceptual framework for the planetary boundary approach, showing the safe operating space, the zone of uncertainty, and the beyond zone of uncertainty. The left figure shows Earth System processes operating at the planetary scale (e.g. the climate system) where the boundary is defined to avoid crossing critical continental or global thresholds. The right figure shows Earth System processes operating at a sub-planetary scale with no known planetary thresholds (e.g. water use or deforestation), but where exceedance of local/regional boundaries can aggregate and add up to affect other Earth System processes and, thus, the general stability of the Earth System. The figure is adapted from Steffen et al. [5].

The quantification of thresholds – known or unknown – is subject to uncertainty due to a lack of perfect scientific knowledge and natural variability in environmental systems. The upper and lower boundaries for this uncertainty generate a zone of uncertainty around each threshold (hatched area in Figure 2). Setting an exact value for a control variable for a certain Earth system process, e.g. maximum ppm $CO_2$ in the Earth's atmosphere to avoid the risks related to Climate Change, is a question of the risk acceptance of human societies. The Planetary Boundaries concept adopts a precautionary approach by defining the Planetary Boundary at the lower bound of the zone of uncertainty to minimize the risk of exceeding thresholds. The Planetary Boundaries, therefore, delimit a "safe operating space for humanity concerning the functioning of the Earth System" given the current scientific knowledge. It is noted that this safe operating space is only present if none of the proposed boundary positions are transgressed.

Table 1 lists the Planetary Boundary data for the Earth system process 'Climate Change'. As can be observed, there are two control variables namely 'Atmospheric

$CO_2$ concentration' measured in ppm $CO_2$ in the atmosphere and 'Energy imbalance at the top of Earth's atmosphere' (i.e. the rate of energy change per unit area of the globe as measured at the top of the atmosphere) measured in $W/m^2$. The difference between the two boundaries is that the latter includes the impact of non-$CO_2$ greenhouse gasses on climate change while the first only covers $CO_2$ and its precursors.

Table 1. Planetary Boundaries for the Earth system 'Climate change'. The Planetary Boundaries are based on the latest version of the Planetary Boundary concept by Steffen et al. [5].

| Climate Change | Control variable | Planetary Boundary |
| --- | --- | --- |
| • Loss of polar ice sheets<br>• Regional climate disruptions<br>• Loss of glacial freshwater supplies<br>• Weakening of carbon sinks | Atmospheric $CO_2$ concentration [ppm] | 350 ppm<br>(350–450 ppm)* |
| | Energy imbalance at the top of Earth's atmosphere [$W/m^2$] | 1 $W/m^2$<br>(1.0-1.5 $W/m^2$)* |

*Zone of uncertainty

## 3  Planetary Boundaries and Life-cycle Assessment

The concept of Planetary Boundaries provides the safe operating space for control variables for the nine Earth system processes but does not link these to indicator scores quantified in life cycle assessments (LCA). Translating the Planetary Boundaries into maximum scores for LCA indicators is therefore necessary if LCA is to support the development of solutions that respect the Planetary Boundaries.

This section describes and applies two different approaches to derive a maximum score for greenhouse gas emissions expressed as annual emissions of $CO_2$-equivalents – an indicator for the LCA impact category 'Climate Change' – that respects the concept of Planetary Boundaries.

*3.1  Method 1: Weighted average of substance-specific contributions*

Bjørn and Hauschild [8] defined maximum score – naming them 'carrying capacities' – as "*the maximum sustained environmental intervention a natural system can withstand without experiencing negative changes in structure or functioning that are difficult or impossible to revert*". The phrase 'sustained environmental intervention' in this definition can be expressed by an LCA indicator score while the latter part of the definition can be regarded as a qualitative description of the ultimate threshold for environmental sustainability. They then use medium or average values of the uncertainty ranges in Rockström et al. [1] as the threshold for environmental sustainability to derive maximum scores for LCA indicators for ten midpoint impact categories[3] commonly included in LCA. The choice to use medium or average values of

---
[3] Impact categories is a categorization system used when quantifying the environmental impact of life-cycle inventories in life-cycle assessments. 'Midpoint impact categories' is the quantification of a benchmark pollutant (indicator) for a certain impact category, e.g. 'kg $CO_2$-equivalent' for the impact category 'Global Warming Potential'. The impact in these

the uncertainty ranges as a threshold for environmental sustainability implies an acceptance of a 50% risk that human-induced impacts on the environment will result in deleterious or even catastrophic environmental changes. This is a different risk philosophy than that of the Planetary Boundaries which recommend the use of the lower bounds of uncertainty to establish a safe operating space for humankind.

However, Bjørn and Hauschild [8] also derived a maximum score for annual emissions of $CO_2$-equivalents based on the Planetary Boundary for radiative forcing (1 W/m$^2$). The translation from radiative forcing to annual emissions of $CO_2$-equivalents was based on the assumption that 1 W/m$^2$ radiative forcing corresponds to an Earth surface temperature increase of 1.06° above pre-industrial levels. Then, the sustained level of emissions of individual greenhouse gasses that would lead to a steady state temperature increase of 1.06° was calculated. The GEOCARB model by Berner and Kothavala [10] implemented by Archer [11] was used to estimate this limit for $CO_2$ by assuming 1) an average pre-industrial (natural) CO2 degassing rate of 7.5E+12 mol/year[4], 2) an equilibrium climate sensitivity (ECS) of 3 °C[5], 3) impact of plants and 4) no change in land area[6], and then identifying the degassing rate of $CO_2$ allowed to reach a steady-state temperature increase of 1.06 °C after 1.95 mills. years. The limits for a range of 21 non-CO2 greenhouse gasses were estimated using the GTP model of Shine et al. [12] and converted into $CO_2$-equivalents using GWP100 global warming potentials for each gas from the 5th IPCC assessment report (AR5) from 2013 [2]. Hereafter, the $CO_2$-equivalents for all greenhouse gasses were weighed concerning their relative fraction of the total $CO_2$-equivalent emissions to the atmosphere in 2008 and 2010 according to Laurent et al. [12] and finally summed up to a safe operating space for the impact category 'Climate Change' expressed in $CO_2$-equivalents: 3.61 gigaton $CO_2$-eq/year.

The calculation by Bjørn and Hauschild [8] was published in 2015 using data from the 5th IPCC assessment report published in 2013 [2]. The 6th IPCC assessment report (AR6) [14] published in 2021 contains an update of the greenhouse gas data used in this calculation; see AR6 supplementary material (7.SM) for details. The next sections report on an update of this specific translation by Bjørn and Hauschild [8] using greenhouse gas data from AR6 and updated global emission data for weighing greenhouse gas-specific contributions to the maximum score. All calculations are available in Appendix 1 (electronic excel spreadsheet).

### 3.1.1 Greenhouse gas lifetime, radiative efficiencies, and GWP100

The original calculation by Bjørn and Hauschild [8] used data on greenhouse gas lifetime, radiative efficiencies, and GWP100 from AR5 to calculate the specific $CO_2$-equivalent contributions of all greenhouse gasses but $CO_2$ itself. Replacing AR5 data with AR6 data reduces the safe operating space by 19.5 % from 3.61 gigatons $CO_2$-eq/year to 2.90 gigaton kg $CO_2$-eq/year.

---

midpoint impact categories affects different endpoint categories such as 'Human Health' and 'Ecosystem Quality'. See e.g. Goedkoop et al. [9] for further details on midpoint and endpoint categories.

[4] Corresponding to an average atmospheric $CO_2$ level of 275 ppm in the pre-industrial period.

[5] ECS is called $T_{2x}$ in this implementation of the GEOCARB model.

[6] Changes in land area changes the net rate of $CO_2$ uptake by weathering reactions.

The reduction is the result of different changes across substance-specific contributions. The realization that the stratospheric-temperature-adjusted radiative efficiency for methane ($CH_4$) in AR6 is increased – with high confidence – by ~25% compared to the values in AR5 is the main driver for the reduction. The effect of this realization on the safe operating space is somewhat counteracted by the revised (reduced) lifetime data for Nitrous Oxide ($N_2O$) and to a minor extent by the revised data for all other greenhouse gasses except for Trichlorofluoromethane (CFC-11).

*3.1.2 Emission data*

The original calculation of the safe operating space by Bjørn and Hauschild [8] included the $CO_2$-equivalents of 22 greenhouse gasses regarded as the most contributing substances to climate change as per Laurent et al. [13][7]. The supplementary material of AR6 (7.SM) provides data on 250 greenhouse gasses; however, only data for the 20 greenhouse gasses that contribute to a 1750–2019 effective radiative forcing (ERF) of at least 0.001 W/m$^2$ are listed in the AR6 main report table 7.5. The updated calculation aligns with this exclusion criterion and, consequently, neglects the ERF from all remaining gasses without further reason. Table 2 shows emission data for the greenhouse gasses included in the original calculation expressed as $CO_2$-equivalents, and the data used in the updated calculation, respectively. The data for the updated calculations are calculated using the newest available emission data and GWP100 from AR6 (see the electronic excel spreadsheet in Appendix 1 for details).

The updated calculation resulted in 2.61 gigatons CO2-eq/year which is 27.6% lower than the original value and 10% lower compared to the calculation with greenhouse gas data updated from AR5 to AR6 data. The main driver for the net reduction remains to be the ~25% increase in the radiative efficiency for methane ($CH_4$) in AR6 compared to AR5.

---

[7] The source does not provide any details on how these were deemed 'most important'.

Table 2. Emissions of various greenhouse gasses (GHG) expressed as $CO_2$-equivalents used in the original and updated

| Original calculation | | | Updated calculation | | |
|---|---|---|---|---|---|
| GHG included | Gigaton $CO_2$-eq in 2010[a] | % | GHG included[b] | Gigaton $CO_2$-eq in 2016[a] | % |
| $CO_2$ | 39.5 | 73.1 | + | 44.9 | 75.9 |
| $CH_4$ | 8.59 | 15.9 | + | 8.2 | 13.9 |
| $N_2O$ | 3.15 | 5.83 | + | 3.0 | 5.12 |
| $SF_6$ | 0.15 | 0.27 | + | 0.20 | 0.33 |
| CFC-11 | 0.34 | 0.63 | + | 0.45 | 0.76 |
| CFC-12 | 0.76 | 1.41 | + | 0.39 | 0.66 |
| CFC-113 | 0.08 | 0.15 | + | 0.05 | 0.08 |
| $CCl_4$ | 0.08 | 0.16 | + | 0.08 | 0.14 |
| HCFC-22 | 0.64 | 1.19 | + | 0.73 | 1.23 |
| HCFC-141b | 0.04 | 0.07 | + | 0.05 | 0.09 |
| HCFC-142b | 0.09 | 0.16 | + | 0.06 | 0.10 |
| $CH_3CCL_3$ | 0.0007 | 0.00 | - | | |
| Halon-1301 | 0.01 | 0.03 | - | | |
| Halon-1211 | 0.01 | 0.02 | - | | |
| Halon-2402 | 0.001 | 0.00 | - | | |
| $CH_3Br$ | 0.00008 | 0.00 | - | | |
| HFC-134a | 0.21 | 0.39 | + | 0.34 | 0.58 |
| HFC-23 | 0.18 | 0.33 | + | 0.18 | 0.31 |
| HFC-143a | 0.08 | 0.14 | + | 0.16 | 0.27 |
| HFC-125 | 0.08 | 0.14 | + | 0.23 | 0.39 |
| HFC-152a | 0.006 | 0.01 | - | - | - |
| HFC-32 | 0.006 | 0.01 | + | 0.03 | 0.05 |
| | | | CFC-115 | 0.01 | 0.02 |
| | | | CFC-114 | 0.00 | 0.00 |
| | | | $C_2F_6$ | 0.01 | 0.02 |
| | | | $CF_4$ | 0.04 | 0.06 |

[a]Emissions are in units of gigatons per year (1 gigaton = 1 000 000 000 ($10^9$) tons). [b]A '+' indicates that the GHG is the same as in Laurent et al. [13], and '-' indicate that the GHG is not included.

### 3.1.3 Uncertainties

This section reports on investigations of how uncertainty in data and different assumptions affect the magnitude of the safe operating space.

Inputs to the GEOCARB model

The GEOCARB model requires an assumption about equilibrium climate sensitivity (ECS), i.e. the increase in Earth's surface temperature per doubling of $CO_2$ content in the atmosphere. The exact value of ECS is debated but seems to converge towards a consensus: AR6 concludes that there is a 90% or more chance (very likely) that the ECS is between 2 °C and 5 °C. This is a significant reduction in uncertainty compared to AR5, which gave a 66% chance (likely) of ECS being between 1.5°C and 4.5°C. ECS was assumed to be 3 °C in the GEOCARB model which resulted in a steady-state temperature increase of 1.06 °C at a steady-state atmospheric $CO_2$ concentration of approx. 350 ppm. This concentration corresponds to the upper limit of the safe operating space for $CO_2$ in the atmosphere according to the concept of Planetary Boundaries. Changing ECS in the GEOCARB model does not affect the safe operating space but corresponds to changing the maximum allowed atmospheric $CO_2$ concentration to keep the temperature increase at 1.06 °C. If ECS is set to 2 °C, then the atmospheric $CO_2$ concentration has nearly converged at a level around 388 ppm after 1.95 mills. years surpassing the upper boundary for the safe operating space. If ECS is set to 5 °C, then the maximum allowed atmospheric $CO_2$ concentration converges at a level around 319 ppm after approx. 1 mill. years.

Whereas ECS in the GEOCARB model does not affect the safe operating space per se, two other input variables do. The input 'Land area' accounts for the net rate of $CO_2$ uptake by weathering reactions due to changes in the land area on the planet relative to today. A land area reduction of 8.5 % relative to the pre-industrial level will eliminate any possibility for $CO_2$ emissions above the pre-industrial level to keep the temperature increase at 1.06 °C. The other GEOCARB variable affecting the calculation of the safe operating space is whether plants on land are allowed to affect the rate of weathering by pumping $CO_2$ into the soil. This mechanism is crucial to preserve in the modelling and reality; if not, atmospheric $CO_2$ will increase to 500 ppm and increase the Earth's surface temperature to 2.7 °C despite keeping $CO_2$ emissions at the pre-industrial level. The magnitude of the safe operating space thus relies on the fulfilment of crucial assumptions about the inputs to the GEOCARB model.

Substance-specific impact

AR6 table 7.15 lists values for the 5% and 95% confidence intervals for lifetime, radiative efficiencies, and GWP100 of five greenhouse gasses included in the updated calculation. AR6 provides no confidence interval for the remaining 15 greenhouse gasses accounted for in the update. Using the values for the 95% confidence interval for all five substances at the same time results in a safe operating space of 2.51 gigaton $CO_2$-eq/year, i.e. ~4% lower than for the 50% confidence interval. AR6 does not provide data with maximum uncertainty but assuming the data is normally distributed, a calculation using 3σ (i.e. confidence of 99.73%) results in a safe operating space of 2.35 gigaton $CO_2$-eq/year, i.e. 10% lower than for the 50% confidence interval.

Emission data

Emission data expressed as $CO_2$-equivalents is calculated using the median/average value of emission spans provided by various sources (see Appendix 1) assuming these values are the most likely. The true value of each emission may be anywhere within these spans thus affecting the size of its $CO_2$-equivalent. Furthermore, the GWP100 values converting the emissions into $CO_2$-equivalents are median values from AR6. No analysis to identify the confidence interval for the safe operating space due to these circumstances is performed. A simple analysis using the extreme (and unlikely) combination of the upper and lower bound of $CO_2$ with the upper and lower bounds of all other greenhouse gasses, respectively, indicates a variation of up to 30%.

## 3.2 Method 2: Characterization factors

Ryberg et al. [15] proposed a methodology that is applicable for assessing sustainability relative to Planetary Boundaries. The methodology uses so-called characterization factors (CFs) to convert life-cycle impact assessments to metrics corresponding to those of the control variables in the Planetary Boundaries framework. This methodology can be reverted to calculate the maximum tolerated indicator score for $CO_2$-equivalent emissions concerning the Planetary Boundary for radiative forcing.

The calculation presented in Appendix 1 (electronic excel spreadsheet) includes the greenhouse gas lifetime, radiative efficiencies, and GWP100 data from AR6 [14] and 2018 data for global GHG emissions (see Appendix 1 for specific sources) to derive a safe operating space for $CO_2$-equivalent emission of 3.79 gigaton $CO_2$-eq/year for the mean values in AR6 across the 20 GHGs included for 2016 in Table 3. Using the 95% confidence interval for AR6 data leads to 3.63 gigaton $CO_2$-eq/year. It is noted that the calculation is based on emissions of 20 GHGs. It should be noted that the magnitude of the annual emission budget expressed in CO2-eq is dependent on the assumed emissions mix. In the default numbers, we apply a current mix of 20 GHGs. If we assume that the mix is reduced to only be comprised of CO2, then the annual emission budget is reduced by 14% if the 95% confidence interval for AR6 data is used. Thus, the composition of the future GHG mix can affect the actual annual emission budget. Due to a lack of knowledge about the future GHG emission mix, we have based the calculation on the GHG emission mix in 2016.

## 3.3 Scenarios for a safe operating space for annual emission of $CO_2$-equivalents

Table 3 summarises the assumptions and results for the various calculations of the maximum score for the annual emissions of $CO_2$-equivalents based on the Planetary Boundary for radiative forcing of 1 $W/m^2$.

Table 3. Scenarios for a safe operating space for the annual emissions of CO2-equivalents based on the Planetary Boundary for radiative forcing of 1 $W/m^2$.

| Scenario | Safe operating space [gigaton $CO_2$-eq/year] | Notes on assumption |
| --- | --- | --- |
| Method 1: Weighted average of substance-specific contributions [8] | | |
| M1-Original | 3.61 | - |
| M1-AR6 | 2.90 | AR5 data replaced with AR6 data |
| M1-AR6+ | 2.61 | AR6 data + only GHG with ERF>0.001 $W/m^2$ included + new emission data |
| M1-AR6+ (95%) | 2.51 | As AR6+ but using the 95% confidence interval of GHG input from AR6 in the GTP model. |
| M1-AR6+ (99.7%) | 2.35 | As AR6+ but using the 99.73% confidence interval of GHG input from AR6 in the GTP model. |
| Method 2: Characterisation factors [15] | | |
| M2-AR6+ | 3.79 | The calculation of CFs is based on data from AR6. Data for 20 global GHG emissions are from 2018. |
| M2-AR6+ (95%) | 3.63 | Using the provided confidence interval of GHG input from AR6 in the GTP model and for GWP-100 leads to the lowest safe operating space |

The following are notes to the calculations:

- The variation in the outcome of the scenarios is up to 38%.
- The outcome of Method 1 and Method 2 is sensitive to the input on annual emission data. No detailed sensitivity analysis was performed but for method 1, a simple sensitivity analysis indicates that this could be ±30% in all scenarios.
- It is noted that the main driver for the difference between the original calculation in Method 1 and the new scenarios is the updated data on methane ($CH_4$) from AR6 which seems to have a ~25% higher radiative efficiency than previously assumed.
- For Method 1, it is noted that the safe operating space in all calculations is derived using methods (GEOCARB [11] and GTP [12]) that have some inherent uncertainties. The safe operating space also relies on the realisation of crucial assumptions about important inputs to the GEOCARB model.

## 4 Conclusion

This paper investigates two different approaches for translating the Planetary Boundary for *Climate Change* into estimations of maximum emission of greenhouse gasses expressed as annual emissions of $CO_2$-equivalents. The outcome of the two different methods varies with ~30% using the same mean or average data for greenhouse gas lifetime, radiative efficiencies, and GWP100 from IPCC report AR6. Changing the data in both methods to the 95% confidence interval reduces the outcome by ~4%. It is noted that the outcome of both estimation methods – for various reasons – are subject to noticeable uncertainty. The estimations presented in this paper may serve as a benchmark for the end goal for proposed regeneration pathways and roadmaps back to the Planetary Boundary for 'Climate Change'.

## 5 Acknowledgment

## 6 References

[1] Rockström, J., W. Steffen, K. Noone, et al. 2009. Planetary boundaries: exploring the safe operating space for humanity. Ecology and Society 14(2): 32. [online] URL: http://www.ecologyandsociety.org/vol14/iss2/art32/

[2] IPCC Fifth Assessment Report (2013). Website: https://www.ipcc.ch/assessment-report/ar5/ (last accessed 16.8.2022)

[3] 2019 Global Status Report for Buildings and Construction Sector. Website: https://www.unep.org/resources/publication/2019-global-status-report-buildings-and-construction-sector (last accessed 16.8.2022)

[4] P.L. Gibbard, M.J. Head, Chapter 30 - The Quaternary Period, Editor(s): Felix M. Gradstein, James G. Ogg, Mark D. Schmitz, Gabi M. Ogg, Geologic Time Scale 2020, 2020, Pages 1217-1255, ISBN 9780128243602, https://doi.org/10.1016/B978-0-12-824360-2.00030-9.

[5]  W. Steffen, J Rockström, K. Richardson et al. 2018. Trajectories of the Earth System in the Anthropocene. Proceedings of the National Academy of Sciences 115 (33) 8252-8259. https://doi.org/10.1073/pnas.1810141115

[6] P. Crutzen. 2002. Geology of mankind. Nature 415 (23). https://doi.org/10.1038/415023a

[7] Steffen W, Richardson K, Rockström J, Cornell SE, Fetzer I, Bennett EM, Biggs R, Carpenter SR, de Vries W, de Wit CA, Folke C, Gerten D, Heinke J, Mace GM, Persson LM, Ramanathan V, Reyers B, Sörlin S. Sustainability. Planetary boundaries: guiding human development on a changing planet. Science. 2015 Feb 3;347(6223):1259855. doi: 10.1126/science.1259855.

[8] Bjørn, A., Hauschild, M.Z. 2015. Introducing carrying capacity-based normalisation in LCA: framework and development of references at midpoint level. Int J Life Cycle Assess 20, 1005–1018. https://doi.org/10.1007/s11367-015-0899-2

[9] Goedkoop, M., Heijungs, R., Huijbregts, M. et al. 2008. ReCiPe: A life cycle impact assessment method which comprises harmonized category indicators at the midpoint and the endpoint level. National Institute for Public Health and the Environment (RIVM). Biithoven, The Netherlands, 2009.

[10] R. A. Berner and Z. Kothavala. 2001. "GEOCARB III: A Revised Model of Atmospheric CO2 over Phanerozoic Time," American Journal of Science 301, 182-204. Doi: 10.2475/ajs.301.2.182.

[11] Archer D., Eby M., Brovkin V. et al. 2009. Annual Review of Earth and Planetary Sciences 2009 37:1, 117-134. http://climatemodels.uchicago.edu/geocarb/

[12] Shine, K.P., Fuglestvedt, J.S., Hailemariam, K. et al. 2005. Alternatives to the Global Warming Potential for Comparing Climate Impacts of Emissions of Greenhouse Gases. Climatic Change 68, 281–302. https://doi.org/10.1007/s10584-005-1146-9

[13] Laurent A, Hauschild MZ, Golsteijn L, et al. 2013. Normalisation factors for environmental, economic and socio-economic indicators. PROSUITE Deliverable 5.2

[14] IPCC Sixth Assessment Report (2021). Website: https://www.ipcc.ch/assessment-report/ar6/ (last accessed 16.8.2022)

[15] Ryberg M.V., Owsianiak M., Richardson K., Hauschild M.Z. 2018. Development of a life-cycle impact assessment methodology linked to the Planetary Boundaries framework. Ecological Indicators 88, 250-262. https://doi.org/10.1016/j.ecolind.2017.12.065